\documentclass[conference]{IEEEtran}
\IEEEoverridecommandlockouts
\usepackage{cite}
\usepackage{amsmath,amssymb,amsfonts}
\usepackage{algorithmic}
\usepackage{graphicx}
\usepackage{textcomp}
\usepackage{xcolor}
\usepackage{subcaption}
\usepackage{enumitem}
\usepackage{multirow}
\usepackage{booktabs}
\usepackage[pagebackref=false,breaklinks=true,colorlinks,bookmarks=false]{hyperref}
\def\BibTeX{{\rm B\kern-.05em{\sc i\kern-.025em b}\kern-.08em
    T\kern-.1667em\lower.7ex\hbox{E}\kern-.125emX}}
\begin{document}

\title{Nine Years of Pediatric Iris Recognition: Evidence for Biometric Permanence

\makeatletter
\newcommand{\linebreakand}{%
  \end{@IEEEauthorhalign}
  \hfill\mbox{}\par
  \mbox{}\hfill\begin{@IEEEauthorhalign}
}
\makeatother

%\thanks{This work is financially supported by the Center for Identification Recognition and Technology (CITeR) and National Science Foundation (NSF) Grant no. \# 1650503}
}

\author{\IEEEauthorblockN{1\textsuperscript{st} Naveenkumar G Venkataswamy}
\IEEEauthorblockA{\textit{Electrical and Computer Engineering} \\
\textit{Clarkson University}\\
New York, USA \\
venkatng@clarkson.edu}

\and
\IEEEauthorblockN{2\textsuperscript{nd} Masudul H Imtiaz}
\IEEEauthorblockA{\textit{Electrical and Computer Engineering} \\
\textit{Clarkson University}\\
New York, USA \\
mimtiaz@clarkson.edu }

\and 
\IEEEauthorblockN{3\textsuperscript{rd} Stephanie Schuckers}
\IEEEauthorblockA{\textit{Electrical and Computer Engineering} \\
\textit{Clarkson University}\\
New York, USA \\
sschucke@clarkson.edu }

}
\IEEEoverridecommandlockouts
\IEEEpubid{\makebox[\columnwidth]{979-8-3503-3607-8/23/\$31.00 ~\copyright2023 IEEE \hfill}
\hspace{\columnsep}\makebox[\columnwidth]{ }}
\maketitle
\IEEEpubidadjcol

\begin{abstract}
Biometric permanence in pediatric populations remains poorly understood despite widespread deployment of iris recognition systems for children in national identity programs such as India's Aadhaar and trusted traveler programs like Canada's NEXUS. This study presents the most comprehensive longitudinal evaluation of pediatric iris recognition to date, analyzing 276 subjects enrolled between ages 4 and 12 and followed for up to nine years through adolescence. Using 18,318 near-infrared iris images acquired semi-annually, we evaluated both commercial (VeriEye) and open-source (OpenIris) recognition systems through linear mixed-effects models that disentangle enrollment age, developmental maturation, and elapsed time while controlling for image quality and physiological factors. False non-match rates remained consistently below 0.5\% across the nine-year observation period for both matchers when using pediatric-calibrated decision thresholds, approaching adult-level performance. However, we reveal significant algorithm-dependent temporal behaviors: VeriEye's apparent performance decline reflects developmental confounding across enrollment cohorts rather than genuine template aging, while OpenIris exhibits modest but genuine temporal aging (0.5 standard deviations over eight years). Image quality and pupil dilation constancy dominated longitudinal performance, with dilation effects reaching 3.0-3.5 standard deviations, substantially exceeding temporal factors. Failures concentrated in 9.4\% of subjects with persistent acquisition challenges rather than accumulating with elapsed time, confirming that acquisition conditions represent the primary performance limitation. These findings provide empirical justification for extending current conservative re-enrollment policies in operational systems, potentially to 10-12 year validity periods for high-quality enrollments at ages 7 or older, and demonstrate that iris recognition remains viable throughout childhood and adolescence when imaging conditions are properly controlled.
\end{abstract}


\section{Introduction and Background}

Biometric permanence, the degree to which a biological trait remains stable enough to support reliable recognition over time, is fundamental to the design and long-term operation of biometric systems. In this study, we operationalize permanence through longitudinal trajectories of genuine match scores and false non-match rates, quantifying how recognition performance evolves with elapsed time, developmental age, and imaging conditions. While biometric permanence is often assumed, few studies have rigorously examined how biological and operational factors affect temporal stability, particularly in pediatric populations where developmental changes introduce variability not observed in adults.

Unlike faces and fingerprints, which exhibit measurable changes due to growth, environmental exposure, and aging~\cite{debayan2017longitudinal,chandaliya2021longitudinal,yoon2015longitudinal,kirchgasser2021plus}, the iris has long been considered comparatively stable~\cite{daugman1994biometric,wildes2002iris}, with iris texture largely formed during gestation. However, empirical validation of this stability assumption in children undergoing rapid physiological development remains sparse. A critical methodological challenge in pediatric longitudinal studies is that elapsed time is inherently confounded with developmental age. Isolating whether performance changes reflect intrinsic template aging, developmental maturation, or enrollment-age cohort effects requires careful statistical decomposition.

Early work demonstrated that accurate iris templates can be captured from toddlers and preschool-aged children~\cite{basak2017multimodal}, though with higher failure-to-acquire rates due to motion, gaze, and focus challenges, with similar findings from field deployments~\cite{masyn2019overcoming}. Despite these challenges, large-scale operational deployments have integrated iris biometrics for children in the absence of longitudinal validation. India's Aadhaar program~\cite{uidai_mbu} permits enrollment from age 5 with mandated re-enrollment at ages 5 and 15, while Canada's NEXUS trusted traveler program~\cite{cbp_children} includes iris biometrics for minors with five-year renewal cycles. These policies lack empirical justification: no longitudinal evidence has established how long iris-based recognition remains reliable during childhood and adolescence.

Longitudinal studies of other modalities provide instructive context. Face recognition exhibits significant degradation over time due to facial growth~\cite{debayan2017longitudinal,guru2020children,best2016automatic,chandaliya2021longitudinal,ricanek2015review,kamble2022machine}, fingerprints show increased intra-subject variability from finger size and skin elasticity changes~\cite{yoon2015longitudinal,kirchgasser2021plus,harvey2018characterization}, and voice biometrics exhibit greater variability due to vocal tract development~\cite{purnapatra2020longitudinal}. These findings underscore that pediatric biometric permanence must be empirically verified for each modality.

Longitudinal evaluations of iris recognition have focused primarily on adults. Studies by Tome et al.~\cite{tome2008effects}, Baker et al.~\cite{baker2009empirical}, Czajka~\cite{czajka2013template}, and the NIST IREX VI report~\cite{grother2015irex} indicate that variability in recognition performance is driven primarily by imaging and physiological conditions, most notably pupil dilation, rather than intrinsic structural changes. However, these findings may not generalize to children, who experience greater physiological and behavioral variability. Prior pediatric iris work~\cite{basak2017multimodal,nelufule2019image} examined capture and quality challenges but did not evaluate longitudinal stability.

Our research group previously analyzed 209 participants aged 4-11 years across three years and found that pupil dilation constancy was a stronger predictor of matching performance than elapsed time~\cite{das2021iris}. We later expanded this work to 230 subjects over 6.5 years, observing consistently low false non-match rates below 1\%~\cite{das2023longitudinal}. While these results demonstrated short- to medium-term stability, they did not capture the full developmental transition through adolescence or disentangle developmental effects from temporal aging. Furthermore, both studies employed only the VeriEye~\cite{VerieyeSDK} commercial matcher, precluding analysis of algorithm-dependent temporal behaviors--a gap the present study addresses by additionally evaluating the open-source OpenIris~\cite{wldiris} system.

The present study provides, to our knowledge, the most extensive longitudinal evidence to date on pediatric iris recognition. We analyze data from 276 subjects enrolled between ages 4 and 12, followed for up to nine years with developmental trajectories extending through age 19. The dataset comprises 18,318 iris images collected under an IRB-approved protocol using a commercial near-infrared sensor. To isolate temporal, developmental, and quality-related effects, we evaluate two iris recognition systems with fundamentally different algorithmic approaches: a commercial matcher (VeriEye SDK~\cite{VerieyeSDK}) and an open-source matcher (OpenIris~\cite{wldiris}). Using linear mixed-effects models with age-period-cohort (APC) parameterizations, we disentangle enrollment age, biological aging, and elapsed time (factors that are linearly dependent in any longitudinal design) while accounting for individual heterogeneity through subject-specific random effects.

Specifically, this study investigates whether iris recognition remains reliable as children mature through adolescence, how enrollment age influences baseline performance and longitudinal stability, how image quality and pupil dilation differences influence recognition outcomes, whether different algorithms exhibit similar temporal behavior in pediatric populations, and what these findings imply for operational practices such as re-enrollment policies.

This work makes three primary contributions. First, we provide empirical evidence that iris recognition remains operationally reliable for up to nine years in children, with false non-match rates consistently below 0.5\% when using pediatric-calibrated decision thresholds. Second, we demonstrate that apparent temporal decline in one commercial matcher reflects developmental confounding across enrollment cohorts rather than intrinsic template aging, while an open-source matcher exhibits modest genuine temporal aging. This matcher-dependent behavior has important implications for algorithm selection in pediatric deployments. Third, we quantify the influence of enrollment age, image quality, and pupil dilation constancy on longitudinal performance. Younger enrollees (ages 4-5) exhibit persistently lower baseline performance compared to older enrollees (ages 8$+$), with effect sizes of 0.5-0.6 standard deviations. Image quality and dilation constancy are dominant predictors, with dilation effects corresponding to 3.0-3.5 standard deviations in one matcher. Overall, our findings suggest that iris recognition remains feasible and stable throughout childhood and adolescence when imaging conditions are well controlled, with practical implications for the design of inclusive biometric systems and evidence-based re-enrollment policies.

Section~\ref{methods} describes the dataset, recognition systems, and modeling framework; Section~\ref{results} presents performance evaluation and statistical results; Section~\ref{discussion} discusses findings and implications; and Section~\ref{conclusion} summarizes contributions.

\section{Methodology}
\label{methods}

This section details the longitudinal dataset and acquisition procedures (Section~\ref{sec:methods_dataset}), iris recognition systems (Section~\ref{sec:methods_systems}), image quality and physiological metrics (Section~\ref{sec:methods_quality}), verification protocol (Section~\ref{method:verification}), performance metrics (Section~\ref{sec:methods_fnmr}), and statistical modeling framework (Section~\ref{sec:methods_stats}).

\subsection{Clarkson Iris Children Dataset}
\label{sec:methods_dataset}
The Clarkson Iris Children Dataset comprises 14 semi-annual collections spanning nine years (2016–2024), with data acquired from local schools in Potsdam, NY, USA. Collections 9–12 were suspended due to COVID-19 restrictions, yielding 14 completed collections (8 pre-pandemic, 6 post-pandemic). Data collection proceeded under Institutional Review Board (IRB) approval with written parental consent and child assent. Initial recruitment targeted children aged 4–12 years; subsequent enrollment prioritized Pre-Kindergarten students (ages 4–5) to maximize longitudinal developmental coverage. Only de-identified metadata (grade level, birth year) were recorded, with unique anonymized identifiers enabling longitudinal tracking across sessions.

Images were acquired using the IrisGuard IG-AD100~\cite{neurotechnology_ig-ad100_2025}, an ISO/IEC 19794-6 compliant NIR sensor, in semi-controlled classroom environments with consistent illumination following standardized protocols, targeting eight images per subject (four per eye) per session. Although the IRIS ID iCAM-T10~\cite{irisid_icamt10_2025} was introduced at Collection 4, we analyze exclusively IG-AD100 captures to maintain sensor consistency and avoid cross-sensor domain shifts that could confound temporal effects.

All 276 enrolled participants attended multiple sessions, with median follow-up duration of 6.5 years (maximum 8.5 years). Session-to-session retention remained high (86.6\% pre-pandemic, 86.2\% post-pandemic), with 165 participants (59.8\%) completing the final collection and 105 (38.0\%) attending over 10 sessions. Enrollment ages ranged from 4–12 years; by the final collection, participant ages spanned 4-19 years. Figure~\ref{fig:subject_participation} illustrates longitudinal participation patterns.

\begin{figure*}[htbp]
    \centering
    \includegraphics[width=0.95\linewidth]{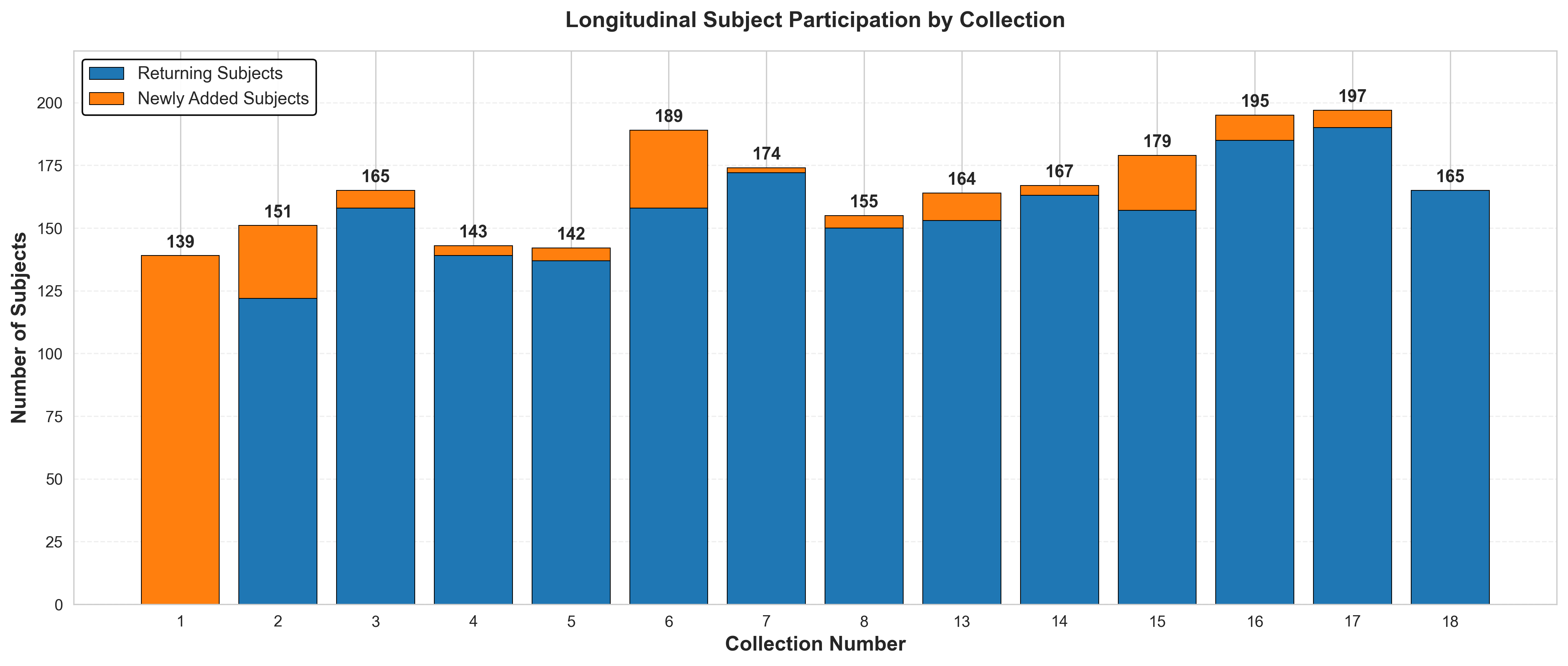}
    \caption{Longitudinal subject participation across collections. Blue bars denote returning participants; orange bars denote new enrollees. Peak participation occurred at Collections 16–17 (n=195–197); the final collection retained 165 subjects (59.8\% of total enrollment).}
    \label{fig:subject_participation}
\end{figure*}

Image acquisition from younger subjects (ages 4–6) presented substantial challenges, as commercial iris sensors are not optimized for pediatric populations. Maintaining gaze fixation, postural stability, and proper device alignment was hindered by physical size constraints. Prevalent imaging artifacts included motion blur, eyelid occlusion, partial eye closure, hair interference, and off-angle gaze deviation.

All images underwent manual review, a total of 97 images (0.52\%) exhibiting severe quality deficiencies were excluded per ISO/IEC 19795-1 guidelines. The final dataset comprised 18,318 valid IG-AD100 images from 276 subjects. Figure~\ref{fig:fta} illustrates representative failure modes.

\begin{figure}[htbp]
    \centering
    \begin{tabular}{ccc}
        \includegraphics[width=0.15\textwidth]{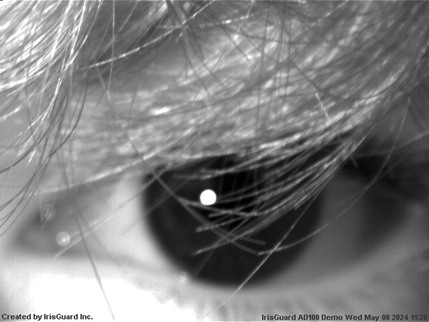} &
        \includegraphics[width=0.15\textwidth]{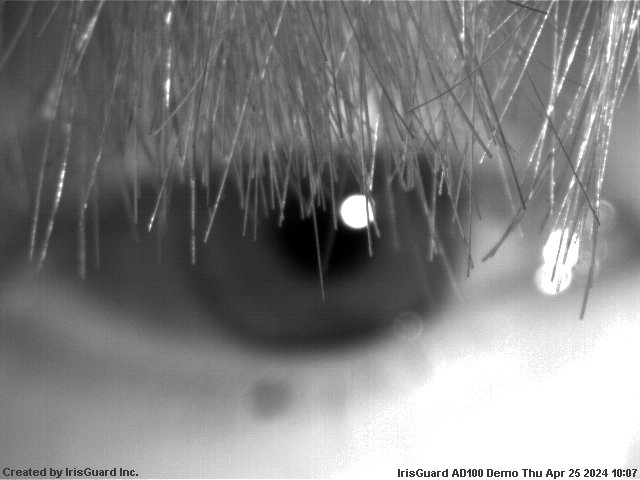} &
        \includegraphics[width=0.15\textwidth]{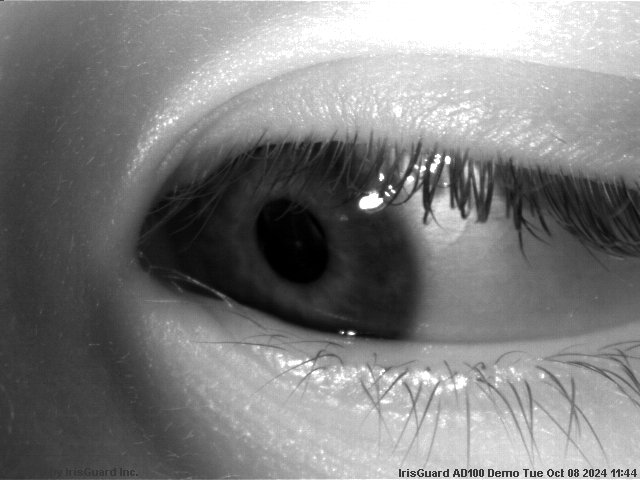}
    \end{tabular}
    \caption{Representative failure modes include off-angle gaze with partial eyelid closure, eyelash occlusion with hair interference, and severe hair occlusion.}
    \label{fig:fta}
\end{figure}

\subsection{Iris Recognition Systems}
\label{sec:methods_systems}

We evaluated two iris recognition systems with fundamentally different design philosophies: a commercial matcher (VeriEye SDK~\cite{VerieyeSDK}) and an open-source matcher (OpenIris~\cite{wldiris}). Both systems processed identical images using the verification protocol described in Section~\ref{method:verification}, enabling algorithm-independent validation of findings.

\subsubsection{VeriEye SDK}
VeriEye 12.4 SDK~\cite{VerieyeSDK} (Neurotechnology) is a commercial iris recognition system compliant with ISO/IEC 29794-6 (iris image quality) and ISO/IEC 19794-6 (iris image format) standards. The processing pipeline comprises automatic segmentation of pupillary and limbic boundaries with occlusion detection, quality assessment generating composite quality scores, template extraction using proprietary feature encoding, and 1:1 comparison producing similarity scores, where higher values indicate greater correspondence. The manufacturer-recommended decision threshold of 36 corresponds to a False Match Rate (FMR) of 0.1\% on adult populations.

\subsubsection{OpenIris Matcher}

OpenIris (v2.0)~\cite{wldiris} is an open-source iris recognition system implementing Daugman's iris code methodology~\cite{daugman1994biometric} with deep learning-based segmentation. The processing pipeline comprises DeepLabv3+ semantic segmentation with ResNet-50 backbone for iris boundary detection, rubber-sheet normalization to compensate for pupil dilation, 2D Gabor wavelet feature extraction producing binary iris codes, and fractional Hamming distance (HD) comparison, where lower distances indicate greater similarity. Match scores range from 0.0 (identical) to 1.0 (maximally dissimilar). The default decision threshold of 0.35 HD is calibrated on adult populations.

\subsubsection{Decision Threshold Calibration}
\label{sec:methods_threshold}
Decision thresholds for both matchers were empirically optimized on the pediatric dataset to balance security and usability. Default thresholds (VeriEye: 36; OpenIris: 0.35 HD) are calibrated on adult populations and assume minimal genuine score variability from high-quality enrollment. However, pediatric iris recognition exhibits elevated genuine score variability due to developmental factors, behavioral challenges during acquisition, and longitudinal template aging across the nine-year observation period.

To maintain equivalent security specifications (FMR $\approx$ 0.1\%) while minimizing false rejections, we computed FMR and FNMR across all decision points using pooled genuine and impostor score distributions from the full dataset. Because our objective is to establish operationally appropriate thresholds for pediatric populations rather than to optimize classifier performance, threshold selection on the evaluation data does not constitute data leakage; the thresholds define the operating point, not the algorithm's discriminative capacity. This optimization produced thresholds of 34 for VeriEye (FMR = 0.11\%, FNMR = 0.16\%) and 0.42 HD for OpenIris (FMR = 0.10\%, FNMR = 0.29\%). Compared to default values, pediatric calibration reduced FNMR by 89\% for OpenIris (2.65\% $\rightarrow$ 0.29\%) while maintaining target security levels. All subsequent analyses employ these calibrated thresholds unless explicitly stated otherwise.

\subsection{Image Quality and Dilation Metrics}
\label{sec:methods_quality}
Longitudinal variability in iris recognition performance is influenced by image quality and physiological factors, particularly pupil dilation~\cite{das2021iris, das2023longitudinal, das2020analysis}. To quantify these effects, we extracted quality scores and dilation measurements using VeriEye's segmentation module and incorporated them as covariates in statistical modeling. For both matchers, quality assessment utilized VeriEye's standardized quality metric to enable direct comparison of quality effects across recognition systems.

\subsubsection{Quality Assessment}
Image quality was assessed using VeriEye's composite quality metric following ISO/IEC 29794-6 guidelines, which generates scalar scores ranging from 0 (unusable) to 100 (optimal). The quality metric integrates focus sharpness, iris-pupil boundary contrast, occlusion extent from eyelids and eyelashes, and gaze deviation from optimal alignment. Quality scores were computed independently for each image and utilized as continuous predictors in longitudinal modeling, with separate scores obtained for gallery images ($Q_{\text{gallery}}$) and probe images ($Q_{\text{probe}}$) in each comparison pair.

\subsubsection{Pupil Dilation Ratio}

Variability in pupil size, influenced by ambient illumination, emotional state, and physiological factors, alters visible iris texture through non-linear deformation and affects recognition accuracy~\cite{das2020analysis}. This effect is particularly pronounced in pediatric populations, where developmental changes influence autonomic pupil reactivity. Pupil dilation ratio ($D$) was defined as the ratio of pupil radius to iris radius:

\begin{equation}
D = \frac{r_{\text{pupil}}}{r_{\text{iris}}}
\end{equation}
where $r_{\text{pupil}}$ and $r_{\text{iris}}$ denote the radii of the pupillary and limbic boundaries, respectively, extracted from VeriEye's automated circular fitting. This normalized measure is dimensionless, theoretically bounded in $[0, 1]$, and invariant to sensor resolution and image scale.

\subsubsection{Dilation Constancy}

Dilation Constancy (DC) quantifies the similarity in pupil dilation between gallery and probe images:

\begin{equation}
DC = 1 - |D_{\text{gallery}} - D_{\text{probe}}|
\end{equation}

Where $D_{\text{gallery}}$ and $D_{\text{probe}}$ represent the dilation ratios of the gallery and probe images, respectively. DC ranges from 0 (maximum dilation difference) to 1 (identical dilation states), with values approaching 1 indicating negligible dilation differences between enrollment and verification. DC was incorporated as a continuous covariate in longitudinal statistical models to assess whether dilation consistency influences recognition stability over time.

\subsection{Verification Protocol and Comparison Design}
\label{method:verification}

This study evaluates both longitudinal template stability (genuine comparisons) and impostor discrimination (impostor comparisons) to provide a comprehensive assessment of pediatric iris recognition performance.

\subsubsection{Genuine Comparison Protocol}

To assess longitudinal stability, we constructed genuine comparisons using a fixed-gallery protocol. For each subject, all iris images from their first attended collection served as gallery templates, with all images from subsequent collections serving as probes. This protocol was applied identically to both VeriEye and OpenIris matchers.

The fixed-gallery design was selected to simulate operational scenarios where enrollment templates remain unchanged without re-enrollment and to establish a consistent temporal reference for each subject, isolating aging effects from session-to-session quality variability. Only participants attending at least two collections contributed to the longitudinal analysis. Each gallery image was compared with all chronologically subsequent images from the same eye, producing temporally separated comparison pairs with enrollment-to-probe intervals spanning up to nine years. This protocol generated 45,927 genuine comparisons (22,609 left eye; 23,318 right eye), with temporal gaps ranging from 6 to 102 months (0.5 to 8.5 years). Left and right eyes were analyzed separately, yielding 551 subject-eye pairs (276 subjects, with some individuals missing data from one eye due to acquisition failures or quality exclusions).

\subsubsection{Impostor Comparison Protocol}

Impostor comparisons were generated to estimate false acceptance rates and evaluate system security under simulated attack scenarios. For reproducibility, the dataset was sorted by subject identifier, eye laterality, and collection number prior to pair generation.

For each eye processed separately, every image in the dataset served as a gallery template. For each gallery image, an impostor pool was constructed consisting of all images from the same eye but from different subjects. From this pool, up to 10 probe images were randomly sampled using a deterministic seed (based on row index) to ensure reproducibility while maintaining computational tractability. This sampling strategy balances comprehensive coverage against the combinatorial explosion of exhaustive pairwise comparisons.

Impostor comparisons were constrained to matching eye laterality (left-to-left, right-to-right only), reflecting realistic attack scenarios where an impostor would attempt verification against enrolled templates of the corresponding eye. Cross-eye comparisons (left-to-right) were excluded as they do not represent plausible operational attacks and would artificially inflate impostor separation.

This protocol generated 138,190 impostor comparison pairs (68,651 left eye; 69,539 right eye) for threshold calibration and false match rate estimation. Combined with genuine comparisons, the evaluation dataset comprises 184,117 total comparisons enabling robust estimation of both FNMR and FMR across the full range of decision thresholds.

\begin{figure*}[htbp]
\centering
\includegraphics[width=0.9\linewidth]{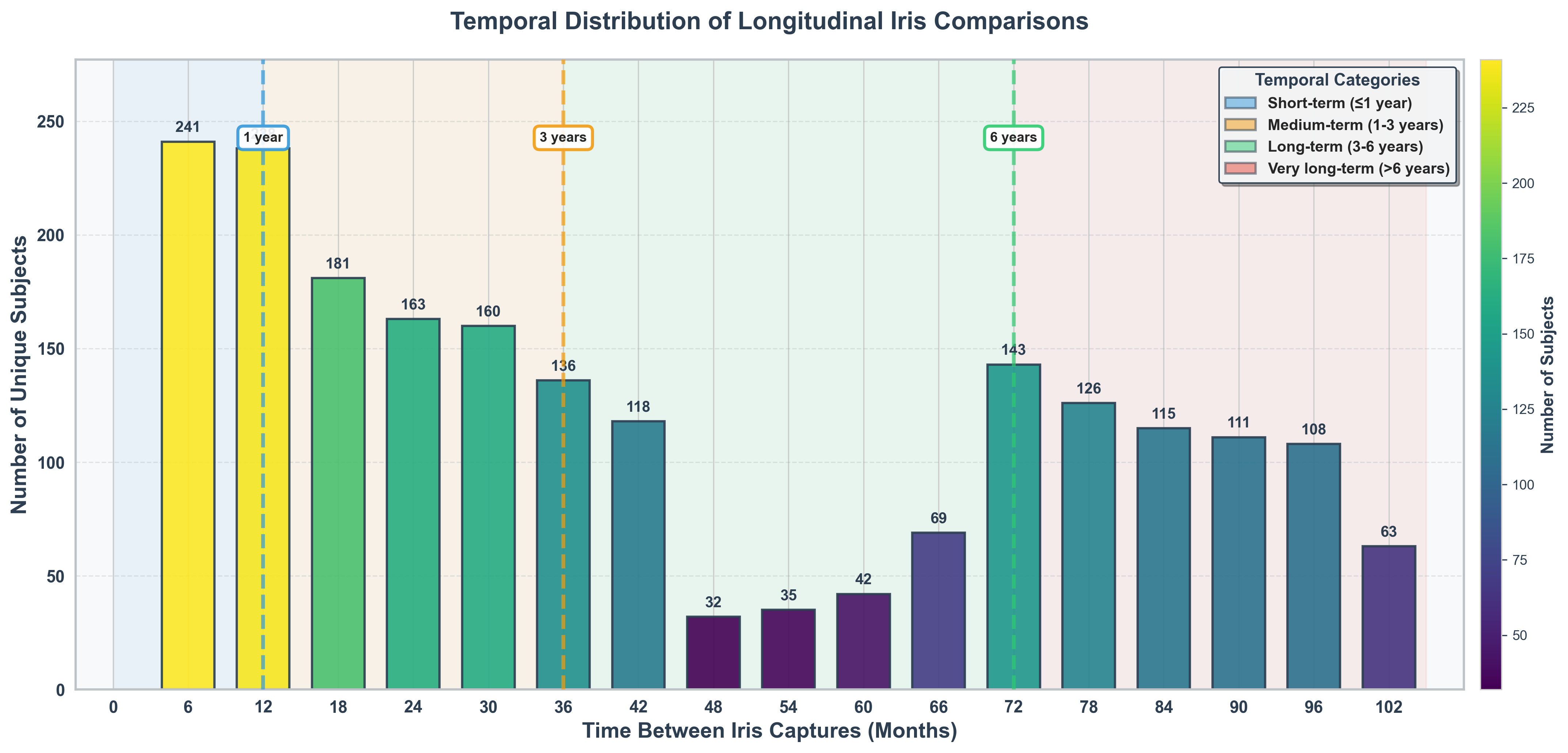}
\caption{Distribution of subjects across temporal intervals. Color zones indicate enrollment-to-probe duration: $\le$1 year, 1–3 years, 3–6 years, and $>$6 years. The reduction at 48–66 months corresponds to COVID-19 suspension (Collections 9–12).}
\label{fig:ad100_month_gap_distribution}
\end{figure*}

Figure~\ref{fig:ad100_month_gap_distribution} illustrates the distribution of subjects contributing genuine comparisons at each temporal interval, stratified by enrollment-to-probe duration. Shorter intervals (6--12 months) demonstrate consistent participation, while extended intervals show sustained longitudinal engagement. A reduction at 48--66 months corresponds to the COVID-19 suspension period (Collections 9--12), after which participation recovered to pre-pandemic levels. The distribution provides adequate statistical power through 96 months (n $>$ 100 subjects), with 63 subjects contributing comparisons at the maximum 102-month interval.

\subsection{Longitudinal Performance Evaluation}
\label{sec:methods_fnmr}
Longitudinal performance was evaluated by computing FNMR at discrete temporal intervals (6, 12, 18, ..., 102 months) corresponding to the semi-annual collection schedule, assessed separately for each matcher (VeriEye and OpenIris) and eye (left and right) to evaluate matcher-specific degradation and potential laterality effects.

A genuine comparison was classified as a false non-match if VeriEye similarity fell below 34 or OpenIris Hamming distance exceeded 0.42 (Section~\ref{sec:methods_threshold}). FNMR at each temporal interval was computed as:

\begin{equation}
\text{FNMR}(t) = \frac{N_{\text{FNM}}(t)}{N_{\text{genuine}}(t)}
\label{eq:fnmr}
\end{equation}

where $N_{\text{FNM}}(t)$ denotes the number of false non-matches and $N_{\text{genuine}}(t)$ represents the total number of genuine comparisons at interval $t$. Each comparison was assigned to a temporal interval based on its enrollment-to-probe duration, rounded to the nearest 6-month increment.

Confidence intervals for FNMR were computed using the Wilson score method~\cite{wilson1927probable}, which provides accurate coverage for proportions near boundaries (0 or 1) without requiring normal approximation assumptions. For intervals with zero observed errors (FNMR = 0), the upper 95\% confidence bound was estimated using the Rule of Three approximation~\cite{mansfield2002best}:

\begin{equation}
\text{FNMR}_{\text{upper}} = \frac{3}{N_{\text{genuine}}(t)}
\label{eq:rule_of_three}
\end{equation}

This approximation provides a conservative upper bound on the true error rate with approximately 95\% confidence when no events are observed in $N$ trials.

\subsection{Statistical Analysis}
\label{sec:methods_stats}
We analyzed longitudinal changes in pediatric iris recognition performance using linear mixed-effects models (LMMs) fitted separately for each matcher (VeriEye, OpenIris) and eye (left, right), treating eyes as independent biometric instances. The analytical objective was explanatory inference, identifying physiological, developmental, and imaging factors associated with variation in matching performance over time.

\subsubsection{Sample and Data Preparation}
All genuine comparisons (same-subject, same-eye pairs across temporal intervals) were included in the analysis. Comparisons missing age information ($<$0.4\% of observations) were excluded. Final sample sizes ranged from 24,997 to 24,004 comparisons per matcher-eye combination (99.6\% retention), representing 274 subjects contributing a mean of 86 comparisons each. Enrollment-to-probe temporal separations spanned 6-102 months. Missingness analyses revealed no systematic association between excluded observations and score distributions. 

\subsubsection{Model Structure}

Matching scores were modeled on their native scales: VeriEye similarity scores ($S$) and OpenIris Hamming distances ($H$). Repeated measurements were nested within subjects and exhibited within-subject correlation (intraclass correlation coefficient $\rho \approx 0.65$), necessitating LMMs with subject-specific random intercepts and random slopes for temporal separation. The general model specification was:

\begin{equation}
    y_{ij} = \beta_0 + \beta_T T_{ij} + \sum_{k \neq T} \beta_k X_{kij} + u_{0i} + u_{1i} T_{ij} + \varepsilon_{ij},
\label{eq:lmm}
\end{equation}

where $y_{ij}$ denotes the matching score for subject $i$ at comparison $j$; $T_{ij}$ represents temporal separation (months) with fixed effect $\beta_T$ (population-average temporal trend) and random slope $u_{1i}$ (subject-specific deviation); $X_{kij}$ denotes additional fixed-effect predictors (age and quality variables); $(u_{0i}, u_{1i})$ are subject-specific random intercepts and slopes assumed jointly normally distributed with covariance matrix $\boldsymbol{\Sigma}$; and $\varepsilon_{ij} \sim \mathcal{N}(0, \sigma^2)$ represents independent residual errors. Parameters were estimated via restricted maximum likelihood (REML) using the statsmodels package (v0.14.0) in Python (v3.10).

\subsubsection{Predictor Variables}

\paragraph{Age and Temporal Variables}
We defined four temporally related variables:
\begin{itemize}
    \item Enrollment age: $A_{\text{gallery}}$ (subject age in years at gallery capture)
    \item Verification age: $A_{\text{probe}}$ (subject age in years at probe capture)
    \item Biological age change: $\Delta A = A_{\text{probe}} - A_{\text{gallery}}$ (years elapsed)
    \item Temporal separation: $T$ (months between gallery and probe captures)
\end{itemize}

In a continuous-time design with exact birth dates and capture timestamps, $\Delta A$ (years) and $T/12$ (years) would be equivalent. However, ages were recorded as integer years and collection intervals were irregular, yielding high but imperfect correlation ($r > 0.99$) and necessitating careful collinearity assessment.

\paragraph{Quality Metrics}
For each comparison, we extracted ISO/IEC compliant quality features for both gallery and probe images:
\begin{itemize}
    \item Overall quality: $Q_{\text{gallery}}, Q_{\text{probe}}$ (0–100 scale)
    \item Usable iris area: $U_{\text{gallery}}, U_{\text{probe}}$
    \item Pupil-boundary circularity: $C_{\text{gallery}}, C_{\text{probe}}$
    \item Pupil-to-iris radius ratio: $R_{\text{gallery}}, R_{\text{probe}}$
    \item Dilation constancy: $DC = 1 - |R_{\text{gallery}} - R_{\text{probe}}|$
\end{itemize}

Fifteen ISO-compliant quality metrics were initially evaluated; the five metrics above were retained for final models based on domain relevance, empirical association with match scores, and collinearity considerations. Variance inflation factors (VIFs) computed for exploratory models including all age/time variables revealed severe collinearity (VIF $>$ 1,000 for $A_{\text{probe}}$) due to the near-identity $A_{\text{probe}} \approx A_{\text{gallery}} + T/12$. After adopting APC consistent two-variable specifications, VIFs remained elevated for some quality metrics (35–170) due to inherent correlations among ISO attributes and gallery-probe analogs. VIFs for temporal separation $T$ (187–193) exceeded conventional thresholds but were deemed acceptable because: (1) $T$ is the primary inferential target; (2) temporal coefficients remained stable across model specifications; and (3) cross-validation confirmed predictive stability.

\subsubsection{APC Parameterization}

Enrollment age ($A_{\text{gallery}}$), verification age ($A_{\text{probe}}$), and temporal separation ($T$) satisfy the approximate linear constraint $A_{\text{probe}} \approx A_{\text{gallery}} + T/12$, creating the well-known APC identification problem~\cite{fosse2019analyzing}. When included simultaneously, their effects become statistically non-identifiable. We therefore compared three alternative two-variable parameterizations:

\begin{enumerate}
    \item $A_{\text{gallery}} + T$ (enrollment age and elapsed time)
    \item $A_{\text{probe}} + T$ (verification age and elapsed time)
    \item $A_{\text{gallery}} + \Delta A$ (enrollment age and biological age change)
\end{enumerate}

Model comparisons based on log-likelihood and Akaike Information Criterion (AIC) revealed matcher- and eye-specific optimal specifications:
\begin{itemize}
    \item VeriEye (both eyes): best fit with $A_{\text{probe}} + T$
    \item OpenIris-Left: best fit with $A_{\text{gallery}} + T$
    \item OpenIris-Right: best fit with $A_{\text{gallery}} + \Delta A$
\end{itemize}
Across all parameterizations, the temporal term remained highly significant ($p < 0.001$), and temporal effect magnitudes were stable. For interpretability and cross-matcher comparability, we adopt $A_{\text{gallery}} + T$ as the primary specification.

\subsubsection{Model Selection and Validation}

Random effects structure was evaluated using likelihood ratio tests comparing models with and without random slopes for temporal separation. Models including random slopes provided significantly better fit across all matcher-eye combinations (LR $\chi^2 = 2{,}217$--$3{,}133$; $p < 0.001$), indicating heterogeneity in individual temporal trajectories. Nested model comparisons confirmed that including temporal separation significantly improved model fit for all groups (LR $\chi^2 = 1{,}652$--$2{,}411$; $p < 0.001$). Five-fold subject-level cross-validation assessed generalization, with out-of-sample $R^2$ values (0.22--0.32) substantially lower than within-sample marginal $R^2$ (0.75--0.77), consistent with high ICC ($\rho \approx 0.65$). Residual diagnostics examined Q-Q plots, residual-versus-fitted plots, and Shapiro-Wilk normality tests, with test statistics ($W = 0.980$--$0.991$) and visual inspection indicating near-normal residual distributions with no systematic patterns suggesting model misspecification.

\subsubsection{Interaction Analysis}
We tested whether temporal trends varied by enrollment age by including an $A_{\text{gallery}} \times T$ interaction term. The interaction was highly significant for VeriEye (both eyes) and OpenIris-Right ($p < 0.001$), but not for OpenIris-Left ($p = 0.055$). For VeriEye, positive interaction coefficients ($\beta = +0.025$ to $+0.112$) indicate that temporal slopes become less negative with increasing enrollment age, reflecting stronger apparent degradation among children enrolled at younger ages (4--7 years) attributable to developmental variability in both behavioral cooperation and iris morphology, whereas children enrolled at older ages ($\ge 10$ years) exhibit more stable trajectories. For OpenIris, the right eye exhibited a small but significant negative interaction ($\beta \approx -0.005$), while the left eye showed no significant interaction. Likelihood ratio tests confirmed that including the interaction improved model fit for VeriEye and OpenIris-Right ($\Delta\text{AIC} = 38$--$240$), but not OpenIris-Left.

\subsubsection{Enrollment Age Group Analysis}
To quantify performance differences by enrollment age while controlling for temporal separation and quality, we discretized enrollment age into four categories: 4–5, 6–7, 8–9, and 10-12 years (reference: 4–5 years). Models included group indicators alongside continuous temporal and quality predictors. Group sample sizes ranged from 1,000 to 11,000 comparisons, providing adequate statistical power for group comparisons.

\subsubsection{Matcher Comparison Analysis}
To test whether VeriEye and OpenIris exhibited different temporal behavior, we constructed combined models within each eye using within-matcher standardized outcomes (z-scores). This standardization removed scale differences, enabling direct comparison of temporal dynamics. A matcher $\times T$ interaction term evaluated whether temporal trajectories diverged between matchers after accounting for baseline performance differences.

\section{Results}
\label{results}

\subsection{Longitudinal Performance Evaluation}
\label{sec:results_fnmr}
Comparison-level FNMR remained consistently below 0.5\% throughout the 9-year observation period for both matchers (Figure~\ref{fig:fnmr_temporal}). VeriEye exhibited a mean FNMR of 0.20\% (left eye: 0.11\%; right eye: 0.30\%), with 11 of 17 left-eye intervals (65\%) exhibiting zero observed errors. OpenIris demonstrated a mean FNMR of 0.33\% (left eye: 0.32\%; right eye: 0.34\%), with comparable performance across eyes. These rates approach manufacturer-reported adult performance under similar FMR constraints, indicating that pediatric populations achieve near-adult accuracy when acquisition quality is controlled.

\begin{figure*}[t]
\centering
\includegraphics[width=0.95\textwidth]{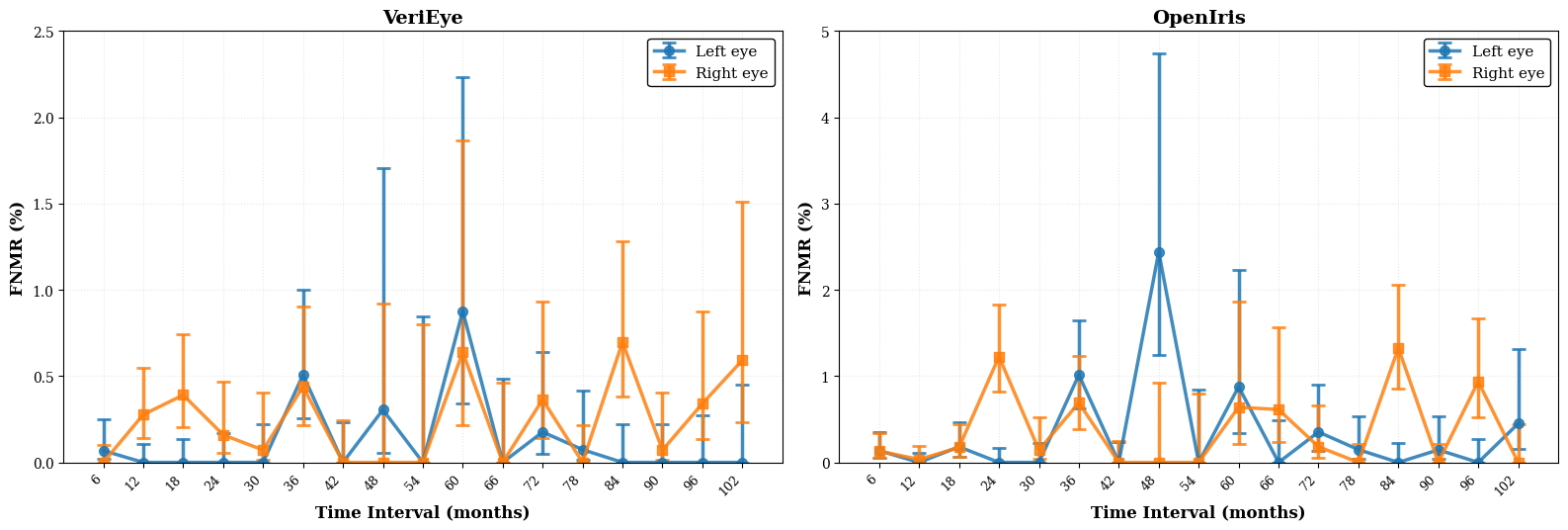}
\caption{Longitudinal FNMR across 6–102 month intervals for VeriEye (left) and OpenIris (right). Both matchers maintained FNMR below 0.5\% for most intervals. Error bars indicate 95\% confidence intervals (Wilson score method; Rule of Three for zero-error intervals). The 48-month elevation (OpenIris-left: 2.42\%, n=330) reflects COVID-19-related sample reduction.}
\label{fig:fnmr_temporal}
\end{figure*}

Temporal analysis revealed gradual increases from early to late intervals. VeriEye increased 2-fold (0.10\% to 0.20\%; 0.10 percentage point absolute change), while OpenIris increased 4.7-fold (0.07\% to 0.33\%; 0.26 percentage point absolute change). Despite these relative increases, absolute FNMR values remained within the 0.5\% operational threshold across the entire 102-month follow-up period. Laterality effects were matcher-dependent: VeriEye exhibited lower FNMR for left eyes (0.11\% vs. 0.30\%), while OpenIris showed minimal eye-specific differences (0.32\% vs. 0.34\%). The transient elevation in OpenIris left-eye FNMR at 48 months (2.42\%, 8/330 failures) coincided with reduced participation during the COVID-19 suspension period (Collections 9-12; see Figure~\ref{fig:subject_participation}), with adjacent intervals exhibiting larger samples and baseline error rates (42 months: 0.12\%, n=812; 54 months: 0.19\%, n=1,024), suggesting the 48-month anomaly reflected sampling variability rather than systematic temporal degradation.

\subsubsection{False Non-Match Analysis}
\label{sec:results_fnm_analysis}

To identify factors associated with recognition failures, we analyzed all genuine comparison pairs where at least one matcher produced a below-threshold score. A total of 164 unique failure pairs were identified, concentrated among 26 subjects (9.4\% of the cohort), indicating that recognition failures are not uniformly distributed but cluster within a subset of individuals with persistent acquisition challenges.

Failure patterns differed substantially between matchers. Of the 164 failure pairs, 20 (12.2\%) were rejected by VeriEye only, 92 (56.1\%) by OpenIris only, and 52 (31.7\%) by both matchers. The predominance of single-matcher failures suggests that VeriEye and OpenIris are sensitive to different image characteristics, a finding confirmed by weak inter-matcher score correlation among failure cases (Pearson $r = 0.131$, $p = 0.095$). This independence implies that the two matchers fail for largely different reasons, supporting the potential value of score-level fusion to reduce false non-match rates.

Contrary to expectations, pupil dilation differences did not significantly predict matcher failures. Correlation analyses between dilation constancy and matcher scores yielded non-significant results across all failure categories (VeriEye-only: $r = -0.073$, $p = 0.758$; OpenIris-only: $r = -0.149$, $p = 0.157$; Both-fail: $r = -0.122$, $p = 0.389$). Instead, image quality metrics emerged as the dominant predictors. Minimum overall quality (the lower of gallery and probe quality scores) showed the strongest association with failures: a threshold of minimum quality $< 45$ captured 96.2\% of both-fail cases, 100\% of VeriEye-only failures, and 63.0\% of OpenIris-only failures. Iris-pupil contrast and sharpness were additionally significant for OpenIris failures ($r = -0.491$ to $-0.568$, $p < 0.0001$), consistent with OpenIris's reliance on fine-scale texture features. Full correlation analyses and threshold evaluations are provided in Supplementary Tables S1--S3.

A notable finding was that 29 of 92 OpenIris-only failures (31.5\%) exhibited pupil boundary circularity values of zero, indicating complete segmentation failure rather than poor matching per se. These segmentation crashes occurred on images with otherwise acceptable quality (mean overall quality = 60.2), suggesting an algorithmic limitation in OpenIris's boundary detection under certain morphological conditions. Excluding segmentation failures, OpenIris-only failures showed quality characteristics similar to the both-fail category.

The 52 both-fail cases represent the most operationally significant failures, as no single matcher could recover these comparisons. These cases exhibited the lowest mean minimum quality (30.6 $\pm$ 8.6), lowest minimum sharpness (79.9 $\pm$ 21.3), and highest dilation variability (delta dilation = 0.107 $\pm$ 0.05) among failure categories. Importantly, both-fail cases were not associated with longer temporal separations (mean 57.1 months) compared to OpenIris-only failures (mean 44.8 months), reinforcing that acquisition conditions rather than elapsed time drive the most severe recognition failures. Descriptive statistics and pairwise comparisons across failure categories are provided in Supplementary Tables S4--S5.

\begin{figure}[t]
\centering
\includegraphics[width=0.95\linewidth]{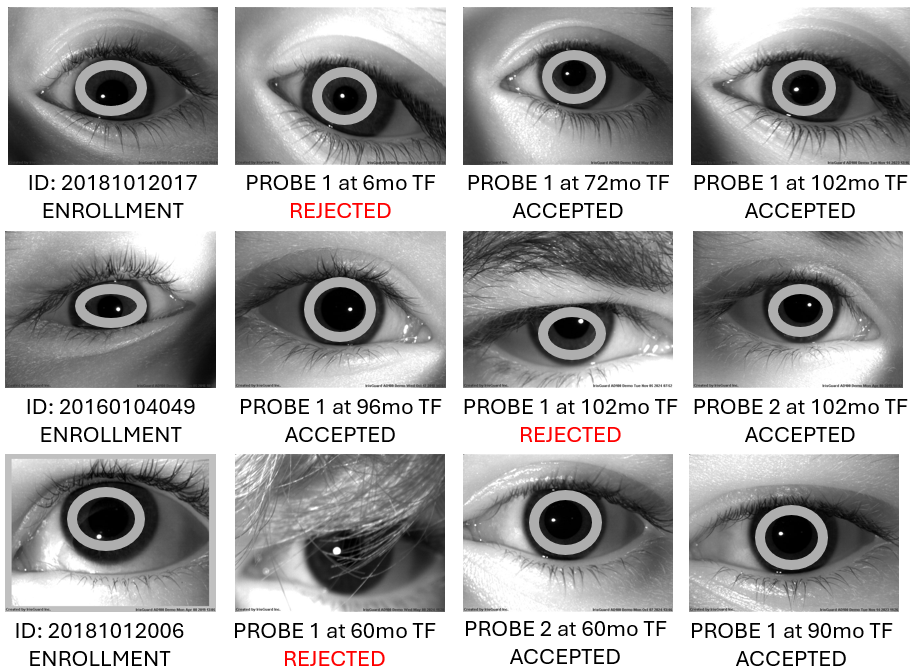}
\caption{Representative false non-matches associated with acquisition challenges (blur, occlusion, reduced quality). Affected subjects later achieved successful verification under improved capture conditions. Note: Iris regions masked for privacy preservation.}
\label{fig:failure_examples}
\end{figure}

Manual inspection of failure cases (Figure~\ref{fig:failure_examples}) confirmed that failures were predominantly associated with acquisition artifacts including motion blur, eyelash and hair occlusion, partial eye closure, and reduced contrast, consistent with the statistical findings. Critically, subjects experiencing failures in one collection frequently achieved successful verification in subsequent collections under improved capture conditions, confirming that failures reflect transient acquisition states rather than permanent template incompatibility.

\subsubsection{False Match Analysis}
\label{sec:results_fm_analysis}

To evaluate system security, we analyzed false acceptance events across 138,190 impostor comparisons. At calibrated decision thresholds (VeriEye $\geq$ 34; OpenIris HD $\leq$ 0.42), VeriEye exhibited a false acceptance rate (FAR) of 0.062\% (85 false accepts) and OpenIris exhibited FAR of 0.070\% (97 false accepts). The overall system achieved equal error rates (EER) ranging from 0.09\% (VeriEye-left) to 0.32\% (OpenIris-right), with area under the ROC curve exceeding 0.999 for all matcher-eye combinations (Figure~\ref{fig:det_curves}). Complete performance metrics including FAR and FRR at multiple operating points are provided in Supplementary Table S6.

Critically, only 3 of the 177 total false accept events (1.7\%) fooled both matchers simultaneously, yielding a combined FAR of just 0.002\% under an AND-rule fusion strategy requiring both matchers to accept. This represents a 98.3\% reduction in false accepts compared to either matcher alone, demonstrating substantial security benefits from multi-algorithm verification. The matchers exhibited complementary failure patterns: 82 impostor pairs (46.3\%) were accepted by VeriEye only, while 92 pairs (52.0\%) were accepted by OpenIris only, confirming that the two systems respond to different iris features.

\begin{table}[t]
\centering
\caption{Critical False Accept Cases: Impostor Pairs Accepted by Both Matchers}
\label{tab:critical_fa}
\small
\begin{tabular}{lccccc}
\toprule
\textbf{Case} & \textbf{Eye} & \textbf{Age (G/P)} & \textbf{VE Score} & \textbf{HD} & \textbf{Min Quality} \\
\midrule
1 & Left & 13 / 7 & 55 & 0.392 & 45 \\
2 & Right & 8 / 18 & 44 & 0.412 & 65 \\
3 & Right & 12 / 17 & 43 & 0.416 & 49 \\
\bottomrule
\end{tabular}
\vspace{1mm}

\footnotesize{G/P = Gallery/Probe subject ages. VE threshold = 34; HD threshold = 0.42.}
\end{table}

The three critical false accept cases where both matchers failed are detailed in Table~\ref{tab:critical_fa}. Case 1 exhibited the highest VeriEye score (55) and lowest Hamming distance (0.392), substantially exceeding acceptance thresholds, suggesting genuine structural similarity between the two unrelated subjects' iris patterns. Notably, this case involved a 6-year age difference (13 vs. 7 years) between gallery and probe subjects. Case 2 involved the largest age gap (10 years, ages 8 and 18) yet produced high-quality images (minimum quality = 65), indicating that developmental stage differences do not preclude false acceptance when iris textures happen to be similar. Case 3 showed marginal scores closer to decision boundaries, representing a less severe but still operationally significant failure. All three cases involved same-eye comparisons with reasonable to good image quality, confirming that these false accepts reflect inherent iris pattern similarity rather than quality-related matching artifacts.

Analysis of factors predicting false acceptance revealed a pattern opposite to false non-matches: image quality showed no significant correlation with false accept occurrence (all $|r| < 0.06$, $p > 0.4$), indicating that quality-based screening cannot prevent false accepts. Subject age emerged as the only statistically significant predictor, with older children ($>$8 years) exhibiting marginally higher confusability ($r = 0.27$, $p < 0.001$). However, this effect was modest, and no quality threshold could effectively screen for false accept risk without substantially impacting genuine acceptance rates. Detailed confusability analysis and subject-level false accept distributions are provided in Supplementary Tables S7--S8.

\begin{figure}[t]
\centering
\includegraphics[width=0.95\linewidth]{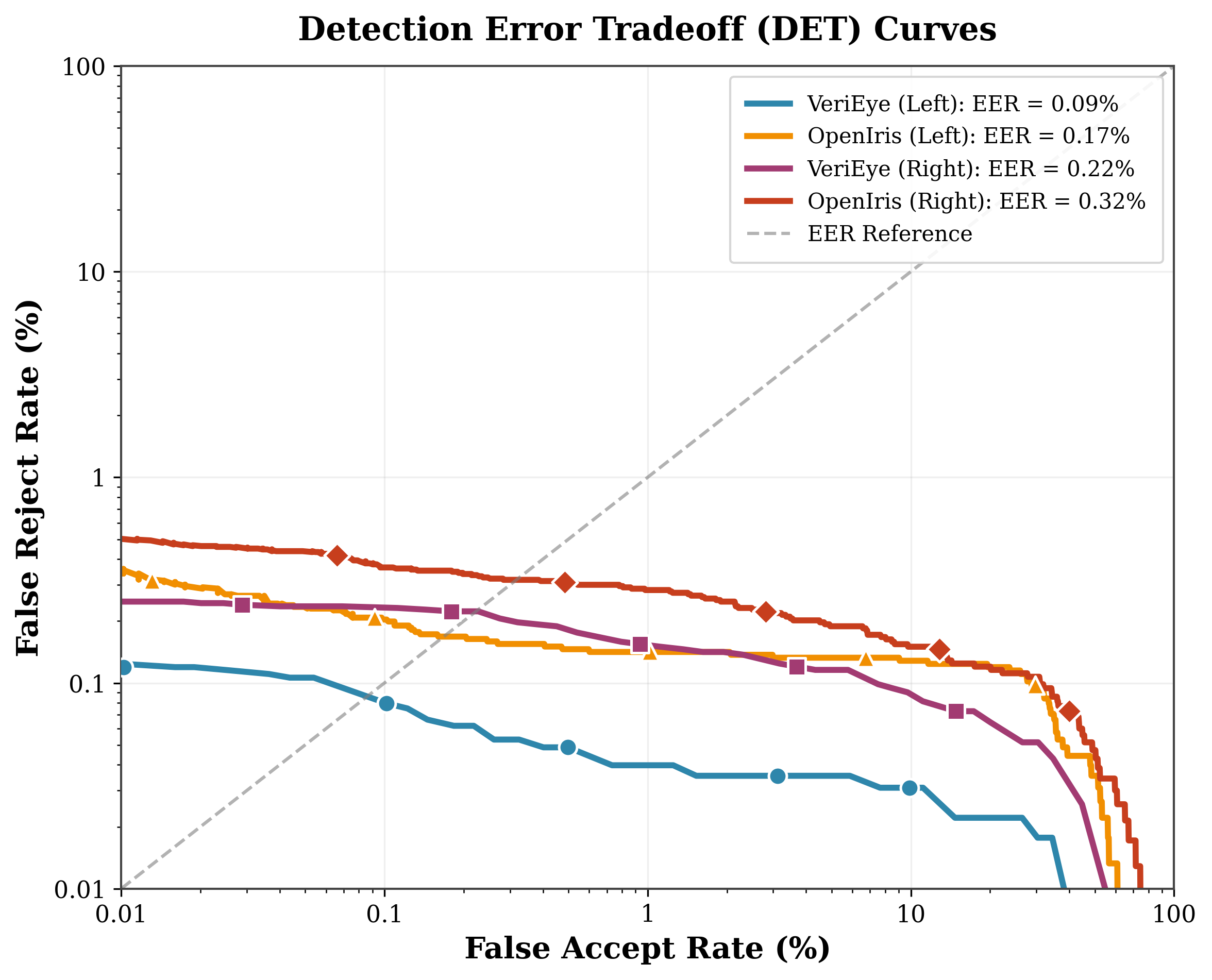}
\caption{Detection Error Tradeoff (DET) curves for VeriEye and OpenIris across both eyes. VeriEye-Left achieved the lowest EER (0.09\%), while all configurations maintained EER below 0.35\%. The separation between curves reflects matcher-specific optimization for the pediatric population.}
\label{fig:det_curves}
\end{figure}

These findings carry direct operational implications. The complementary failure patterns of VeriEye and OpenIris, with 98.3\% of false accepts detectable by at least one matcher, strongly support multi-algorithm fusion for high-security pediatric deployments. An AND-rule requiring both matchers to accept would reduce FAR from 0.13\% (either matcher) to 0.002\% while only modestly increasing FNMR, as the matchers also exhibit complementary patterns for genuine match failures.

\subsection{Sample Characteristics and Multivariate Modeling Rationale}
\label{sec:results_lmm}
The descriptive FNMR analysis revealed gradual temporal increases, matcher-dependent laterality effects, and failure concentration among subjects with persistent acquisition difficulties. However, observational FNMR conflates temporal separation, developmental age changes, enrollment cohort effects, and image quality variability. To disentangle these factors, we employed linear mixed-effects models controlling for quality covariates and accounting for subject-level heterogeneity.

Intraclass correlation coefficients were high (0.649-0.655), indicating that approximately 65\% of performance variance is attributable to stable between-subject differences, with 35\% reflecting within-subject variation from temporal changes and quality factors. Quality metrics remained stable across collections (all trend tests $p > 0.05$), confirming consistent acquisition protocols throughout the study period.

\subsection{Model Structure and Selection}
\label{sec:model_structure} 
Likelihood ratio tests strongly favored random-intercept plus random-slope LMMs over random-intercept-only models for all four groups ($df = 2$; all $p < 0.001$), indicating heterogeneity in subject-specific temporal trajectories. Intraclass correlation coefficients were uniformly high (ICC = 0.649-0.655), with approximately 65\% of performance variance attributable to stable between-subject differences and 35\% to within-subject variation. This high ICC indicates that individual baseline performance is the dominant source of variability, with temporal changes and quality factors contributing comparatively smaller proportions.

To address the near-linear dependency $A_{\text{probe}} \approx A_{\text{gallery}} + T/12$, three identifiable APC parameterizations were compared: (1) $A_{\text{gallery}} + T$, (2) $A_{\text{probe}} + T$, and (3) $A_{\text{gallery}} + \Delta A$, where $\Delta A$ is biological age change. VeriEye models showed marginally better fit under the $A_{\text{gallery}} + \Delta A$ parameterization ($\Delta\text{AIC} = 4$--$20$), whereas OpenIris models favored $A_{\text{gallery}} + T$ for the left eye and $A_{\text{gallery}} + \Delta A$ for the right eye (all $\Delta\text{AIC}$ within 4--20, indicating statistically equivalent fits). Across all APC parameterizations, the temporal term remained highly significant ($p < 0.001$) and directionally consistent, confirming that conclusions regarding temporal effects are robust to APC specification.

\begin{table*}[!t]
\centering
\caption{Fixed-Effect Estimates from Linear Mixed-Effects Models ($A_{\text{gallery}} + T$ Parameterization)}
\label{tab:fixed_effects}
\setlength{\tabcolsep}{3pt} % tighten column spacing
\scriptsize % or \small if you prefer
\resizebox{\textwidth}{!}{%
\begin{tabular}{lcccccccc}
\toprule
& \multicolumn{2}{c}{\textbf{VeriEye-Left}} 
& \multicolumn{2}{c}{\textbf{VeriEye-Right}} 
& \multicolumn{2}{c}{\textbf{OpenIris-Left}} 
& \multicolumn{2}{c}{\textbf{OpenIris-Right}} \\
\cmidrule(lr){2-3}
\cmidrule(lr){4-5}
\cmidrule(lr){6-7}
\cmidrule(lr){8-9}
\textbf{Predictor} 
& $\beta$ (SE) & \textit{p}
& $\beta$ (SE) & \textit{p}
& $\beta$ (SE) & \textit{p}
& $\beta$ (SE) & \textit{p} \\
\midrule
Enrollment age ($A_\text{gallery}$)           
& \textbf{7.58} (2.92) & 0.002
& \textbf{5.23} (2.92) & 0.032
& \textbf{-0.0036} (0.0011) & $<\!0.001$
& \textbf{-0.0024} (0.0012) & 0.014 \\

Temporal gap ($T$)                        
& \textbf{-0.60} (0.02) & $<\!0.001$
& \textbf{-0.68} (0.02) & $<\!0.001$
& \textbf{0.00024} (0.00005) & $<\!0.001$
& \textbf{0.00029} (0.00006) & $<\!0.001$ \\

Enrollment quality ($Q_\text{gallery}$)        
& \textbf{1.59} (0.17) & $<\!0.001$
& \textbf{0.86} (0.19) & $<\!0.001$
& \textbf{-0.00037} (0.00007) & $<\!0.001$
& 0.00014 (0.00008) & 0.064 \\

Verification quality ($Q_\text{probe}$)      
& \textbf{1.19} (0.05) & $<\!0.001$
& \textbf{0.52} (0.06) & $<\!0.001$
& \textbf{-0.00032} (0.00002) & $<\!0.001$
& \textbf{-0.00017} (0.00002) & $<\!0.001$ \\

Enrollment usable area ($U_\text{gallery}$)    
& \textbf{-0.83} (0.21) & $<\!0.001$
& \textbf{3.23} (0.25) & $<\!0.001$
& 0.00010 (0.00008) & 0.19
& \textbf{-0.0020} (0.0001) & $<\!0.001$ \\

Verification usable area ($U_\text{probe}$)  
& \textbf{1.99} (0.07) & $<\!0.001$
& \textbf{2.44} (0.07) & $<\!0.001$
& \textbf{-0.0010} (0.0000) & $<\!0.001$
& \textbf{-0.0011} (0.0000) & $<\!0.001$ \\

Dilation constancy ($DC$)        
& \textbf{438.6} (8.5) & $<\!0.001$
& \textbf{359.8} (8.8) & $<\!0.001$
& \textbf{-0.124} (0.003) & $<\!0.001$
& \textbf{-0.104} (0.003) & $<\!0.001$ \\

Enrollment circularity ($C_\text{gallery}$)    
& \textbf{3.62} (0.19) & $<\!0.001$
& \textbf{2.04} (0.17) & $<\!0.001$
& \textbf{-0.0020} (0.0001) & $<\!0.001$
& \textbf{-0.0015} (0.0001) & $<\!0.001$ \\

Verification circularity ($C_\text{probe}$)  
& \textbf{1.18} (0.07) & $<\!0.001$
& \textbf{2.18} (0.07) & $<\!0.001$
& \textbf{-0.00091} (0.00003) & $<\!0.001$
& \textbf{-0.00105} (0.00003) & $<\!0.001$ \\
\bottomrule
\end{tabular}%
}
\end{table*}

For consistency and interpretability across all matcher-eye combinations, Table~\ref{tab:fixed_effects} reports fixed-effect results from the common $A_{\text{gallery}} + T$ parameterization. However, to demonstrate that VeriEye's temporal patterns are substantively driven by developmental factors rather than elapsed time per se, Section~\ref{sec:verieye_results} additionally reports results from the $A_{\text{gallery}} + \Delta A$ parameterization, which provides complementary insight into biological aging versus enrollment cohort effects.

Under the primary $A_{\text{gallery}} + T$ specification, marginal $R^2$ values (Nakagawa approximation) ranged from 0.748 to 0.770 across matcher-eye groups, indicating that fixed effects (age, time, quality covariates) explain approximately 75\% of within-subject variance after accounting for individual baseline differences. Nested model tests comparing models with and without temporal terms confirmed that time significantly improved model fit for all groups (all $p < 0.001$). However, as detailed in Sections~\ref{sec:verieye_results} and~\ref{sec:openiris_results}, the nature and interpretation of temporal effects differed by matcher: VeriEye's apparent temporal decline reflected developmental confounding, whereas OpenIris exhibited genuine temporal aging independent of developmental factors.

\subsection{Fixed Effects: Age, Time, and Quality}

Table~\ref{tab:fixed_effects} reports fixed-effect estimates for all predictors from the primary $A_{\text{gallery}} + T$ models. The temporal effects in these models revealed different patterns across matchers: VeriEye's apparent temporal decline was fully explained by developmental confounding, whereas OpenIris exhibited genuine temporal aging independent of developmental factors. We present these findings separately below.

\subsubsection{VeriEye: Developmental Confounding, Not Template Aging}
\label{sec:verieye_results}
In the primary $A_{\text{gallery}} + T$ specification (Table~\ref{tab:fixed_effects}), enrollment age was positively associated with VeriEye match scores ($\beta = 5.23$-$7.58$, $p \le 0.032$), indicating that older children produced more stable templates at enrollment. Temporal separation showed negative associations ($\beta = -0.60$ to $-0.68$, $p < 0.001$), consistent with apparent performance decline over time.

However, alternative APC parameterizations and diagnostic specifications revealed that this apparent temporal degradation was fully explained by developmental factors. In the $A_{\text{gallery}} + \Delta A$ parameterization, biological age change $\Delta A$ was negatively associated with VeriEye scores (Left: $\beta = -7.29$, 95\% CI [$-11.58$, $-3.00$], $p < 0.001$; Right: $\beta = -8.17$, 95\% CI [$-12.51$, $-3.83$], $p < 0.001$), while enrollment age remained positively associated (Left: $\beta = 7.20$, 95\% CI [2.50, 11.90], $p = 0.003$; Right: $\beta = 4.86$, 95\% CI [0.13, 9.59], $p = 0.047$).

In a supplementary overidentified diagnostic model simultaneously including $A_{\text{gallery}}$, $A_{\text{probe}}$, and $T$ (which violates statistical identifiability but provides interpretive insight), the temporal coefficient reversed sign and became non-significant (e.g., Right: $\beta = -0.16$, $p = 0.24$), while both age terms remained highly significant ($p < 0.001$). Adding age variables attenuated the temporal coefficient by over 100\%, indicating that apparent temporal decline reflects developmental confounding across enrollment cohorts rather than intrinsic template aging.

\subsubsection{OpenIris: Modest but Genuine Temporal Aging}
\label{sec:openiris_results}

OpenIris exhibited a distinct pattern. In the primary $A_{\text{gallery}} + T$ models (Table~\ref{tab:fixed_effects}), the temporal coefficient remained significant after controlling for enrollment age (Left: $\beta = 0.00024$, 95\% CI [$0.00014$, $0.00034$]; Right: $\beta = 0.00029$, 95\% CI [$0.00018$, $0.00040$]; both $p < 0.001$), corresponding to an increase of approximately 0.023--0.028 HD units over an 8-year separation (96 months), or approximately 0.5 standard deviations. Despite statistical significance, this magnitude corresponds to the modest FNMR increases observed in Section~\ref{sec:results_fnmr} (0.07\% to 0.33\% over 102 months), which remained within operationally acceptable thresholds ($<$0.5\%). Thus, while OpenIris exhibits genuine temporal aging, the effect is operationally manageable over the 9-year observation period when image quality is controlled.

\begin{figure*}[!t]
\centering
\includegraphics[width=\textwidth]{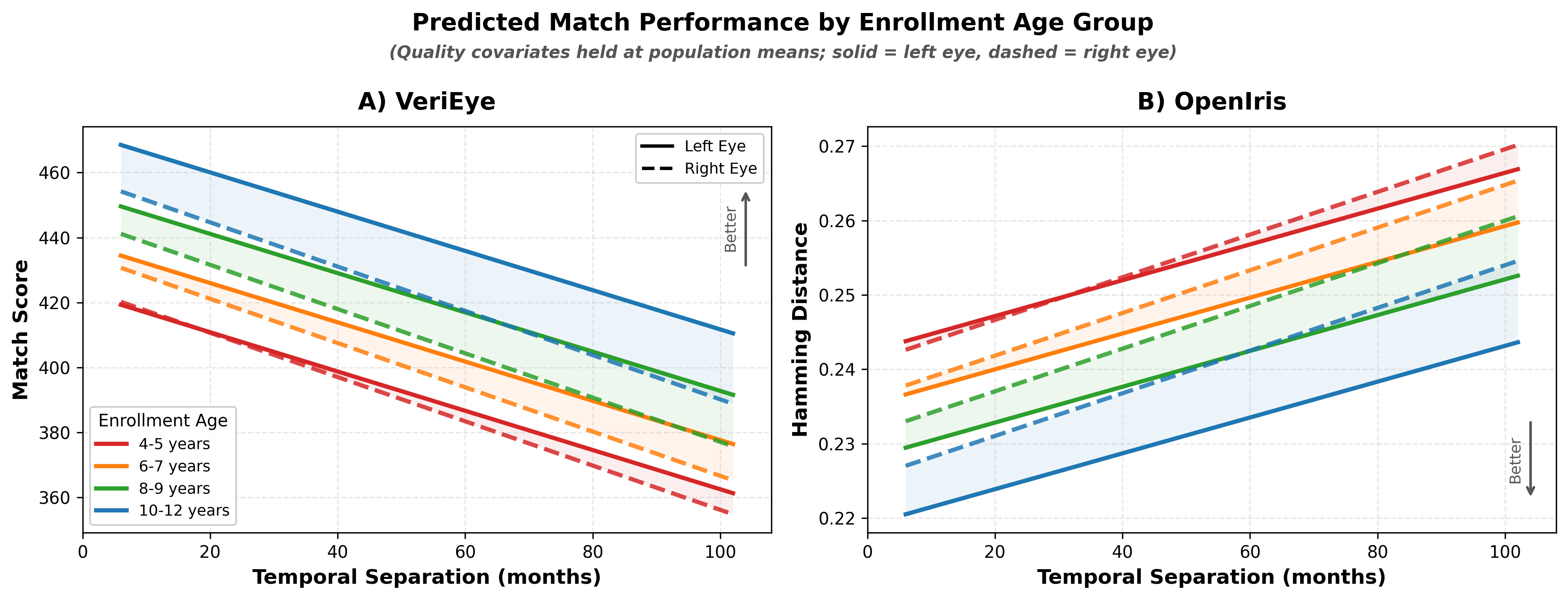}
\caption{Predicted match performance by enrollment age group for (A)~VeriEye and (B)~OpenIris, with quality covariates at population means (solid = left eye; dashed = right eye). Parallel trajectories with baseline offsets (VeriEye) indicate developmental confounding, whereas uniform upward slopes (OpenIris) confirm genuine temporal aging.}
\label{fig:trajectories}
\end{figure*}

Age effects for OpenIris were weak and inconsistent. Enrollment age showed small negative associations ($\beta = -0.0024$ to $-0.0036$, $p = 0.01$--$0.02$), while verification age coefficients were near zero and non-significant in the $A_{\text{probe}} + T$ parameterization ($\beta = 0.0004$--$0.0005$, $p > 0.45$), indicating minimal developmental confounding. The temporal effect persisted across all three APC parameterizations, providing evidence for a modest but genuine temporal aging effect in OpenIris scores, largely independent of developmental factors (Figure~\ref{fig:trajectories}B).

\subsection{Quality and Physiological Predictors}
Beyond age and temporal effects, three predictor classes exerted consistent, substantive influence on matching performance across all models (Table~\ref{tab:fixed_effects}).

Image quality showed strong associations with performance across both matchers. Enrollment and verification quality exhibited significant effects (VeriEye: $\beta = 0.86$--1.59; OpenIris: $\beta = -0.00014$ to $-0.00037$; all $p < 0.001$), with enrollment quality generally showing larger effects than verification quality, consistent with the permanence of enrollment template characteristics. Usable iris area was similarly influential (VeriEye: $\beta = -0.83$ to 3.23; OpenIris: $\beta = -0.0010$ to $-0.0020$; all $p < 0.001$).

Dilation constancy emerged as the strongest individual predictor for VeriEye performance, with coefficients reflecting the combined influence of pupil size mismatch and its correlated effects on iris texture geometry, rubber-sheet mapping distortion, occlusion patterns, and usable area ($\beta = 360$--439, $SE = 8.5$--8.8, both $p < 0.001$). A perfect dilation match ($DC = 1.0$) versus complete mismatch ($DC = 0.0$) predicted 360--439 point score differences, corresponding to approximately 3.0--3.5 standard deviations. OpenIris exhibited smaller but significant dilation effects ($\beta = -0.104$ to $-0.124$, $p < 0.001$), representing approximately 0.1 Hamming distance units over the full dilation range.

Pupil-boundary circularity showed positive associations with VeriEye scores ($\beta = 1.2$--3.6, all $p < 0.001$) and negative associations with OpenIris Hamming distances ($\beta = -0.0009$ to $-0.0020$, all $p < 0.001$), indicating that well-formed circular pupils facilitated template stability in both systems. Standardized effect sizes (Figure~\ref{fig:forest}) confirm that image quality factors, particularly dilation constancy ($\beta = 0.17$--$0.21$) and pupil circularity ($\beta = 0.12$--$0.27$), exceed the magnitude of temporal separation effects ($\beta = -0.16$ to $-0.18$).

\begin{figure}[!t]
\centering
\includegraphics[width=\columnwidth]{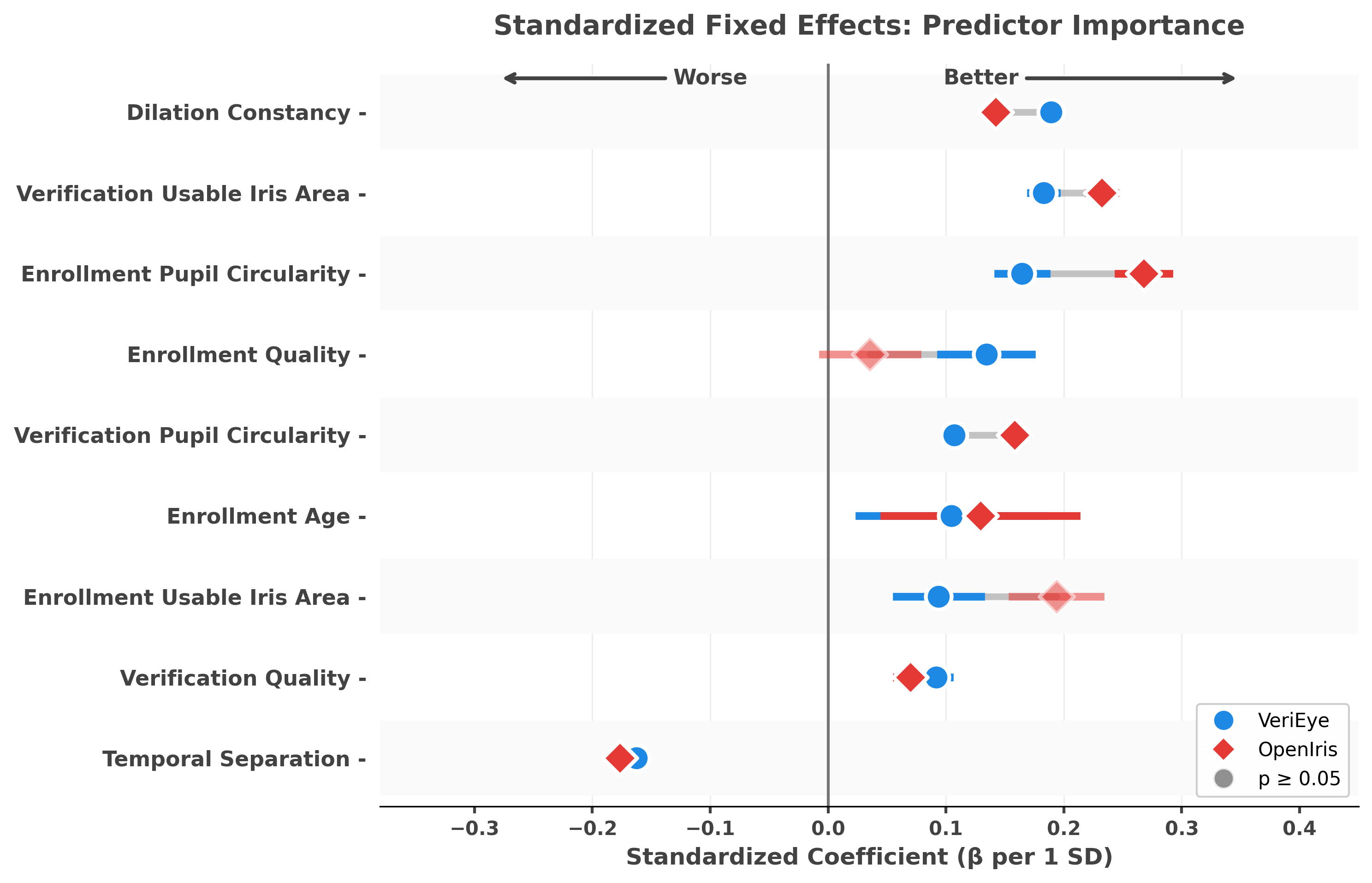}
\caption{Standardized fixed effects by matcher (positive = better performance; faded = p $\geq$ 0.05). Quality factors show larger effects than temporal separation, indicating acquisition conditions dominate over template aging.}
\label{fig:forest}
\end{figure}

\subsection{Age-by-Time Interactions}
Enrollment age significantly moderated temporal patterns for VeriEye (Left: $\beta = 0.025$, 95\% CI [$0.008$, $0.042$]; Right: $\beta = 0.112$, 95\% CI [$0.090$, $0.134$]; both $p < 0.001$). Positive interaction coefficients indicate that temporal slopes become less steep with increasing enrollment age. For example, in the right eye, a child enrolled at age 10 exhibited an estimated 0.67 points per month slower rate of apparent decline than one enrolled at age 4, consistent with cohort-based performance differences rather than uniform aging-related decline (Figure~\ref{fig:trajectories}A).

For OpenIris, the interaction was non-significant for the left eye ($\beta = -0.000013$, $p = 0.06$) but significant for the right eye ($\beta = -0.000036$, 95\% CI [$-0.000051$, $-0.000021$], $p < 0.001$), with substantially smaller magnitude than VeriEye (translating to only 0.00022 HD units per month difference over a 6-year enrollment age span). Likelihood ratio tests confirmed that including the interaction improved model fit for VeriEye and OpenIris-Right ($\Delta\text{AIC} = 38$-$240$), but not OpenIris-Left.

\subsection{Enrollment Age Group Effects}

To quantify developmental differences in operationally interpretable terms, subjects were stratified into four enrollment age groups with the youngest group (4-5 years) serving as the reference category. Children enrolled at 10-12 years significantly outperformed those enrolled at 4--5 years for both matchers, with effect sizes of 0.5--0.6 SD (VeriEye: $+63$-$71$ points, $p \le 0.013$; OpenIris: $-0.029$ to $-0.030$ HD, $p \le 0.003$; Table~\ref{tab:age_groups}). The 8-9 year group showed intermediate effects, while the 6-7 year group exhibited minimal differences from the reference group. These age-stratified results indicate that enrollment at ages $\ge$8 years yields more stable baseline performance, consistent with developmental maturation reducing acquisition challenges and template variability.

\begin{table*}[!t]
\centering
\caption{Enrollment Age Group Effects on Longitudinal Match Performance (Reference: 4-5 Years)}
\label{tab:age_groups}
\small
\begin{tabular}{lcccc}
\toprule
\textbf{Matcher-Eye} & \textbf{Age Group} & \textbf{$\beta$ [95\% CI]} & \textbf{p} & \textbf{d} \\
\midrule
VeriEye-Left   & 6-7 years     & 10.35 [-14.1, 34.8]      & 0.400  & 0.08 \\
               & 8-9 years     & \textbf{35.26} [7.1, 63.4]       & 0.015  & 0.28 \\
               & 10-12 years & \textbf{71.16} [23.2, 119.1]     & 0.004  & 0.57 \\
\midrule
VeriEye-Right  & 6-7 years     & 8.02 [-14.7, 30.7]       & 0.521  & 0.06 \\
               & 8-9 years     & 15.27 [-12.1, 42.6]      & 0.277  & 0.12 \\
               & 10-12 years & \textbf{62.92} [13.7, 112.2]     & 0.013  & 0.50 \\
\midrule

OpenIris-Left  & 6-7 years     & -0.0052 [-0.0149, 0.0044]  & 0.272  & 0.11 \\
               & 8-9 years     & \textbf{-0.0188} [-0.0293, -0.0084] & $<$0.001 & 0.39 \\
               & 10-12 years & \textbf{-0.0286} [-0.0393, -0.0179] & 0.002  & 0.60 \\
\midrule
OpenIris-Right & 6-7 years     & -0.0051 [-0.0143, 0.0041]  & 0.306  & 0.11 \\
               & 8-9 years     & -0.0077 [-0.0188, 0.0035]  & 0.171  & 0.16 \\
               & 10-12 years & \textbf{-0.0301} [-0.0409, -0.0192] & 0.003  & 0.63 \\
\bottomrule
\end{tabular}
\end{table*}

\subsection{Matcher Comparison}
Combined z-score standardized models revealed significant matcher-by-time interactions in both eyes (Left: $\beta = -0.0085$, 95\% CI [$-0.0093$, $-0.0077$]; Right: $\beta = -0.0073$, 95\% CI [$-0.0081$, $-0.0065$]; both $p < 0.001$), confirming that VeriEye and OpenIris exhibit statistically distinct temporal trajectories after controlling for age, quality, and dilation. The negative interaction indicates divergent patterns: VeriEye's apparent decline reflects developmental confounding (Section~\ref{sec:verieye_results}), whereas OpenIris exhibits genuine temporal aging (Section~\ref{sec:openiris_results}).

\subsection{Random Effects and Individual Heterogeneity}

Random intercepts showed large standard deviations (VeriEye: $\sigma_{\text{intercept}} \approx 83$ points; OpenIris: $\sigma_{\text{intercept}} \approx 0.03$ HD units), indicating that individuals vary by approximately $\pm$166 VeriEye points or $\pm$0.06 HD units ($\pm$2 SD) around the population mean, independent of measured covariates. Random slopes also exhibited meaningful heterogeneity (VeriEye: $\sigma_{\text{slope}} = 0.8$--1.2 points/month; OpenIris: $\sigma_{\text{slope}} = 0.0001$--0.0002 HD/month), corresponding to approximately $\pm$2.0 points/month (VeriEye) or $\pm$0.0004 HD/month (OpenIris) variability in individual temporal trajectories. The correlation between random intercepts and slopes was weak to moderate ($\rho = -0.2$ to $+0.3$), indicating that individuals with higher baseline performance do not systematically exhibit steeper or shallower temporal trajectories.

\subsection{Model Validation and Generalization}
\label{sec:validation}

Residual diagnostics confirmed no major violations of model assumptions. Q-Q plots showed near-linear patterns, residual-versus-fitted plots revealed no systematic patterns or heteroscedasticity, and Shapiro-Wilk tests yielded $W = 0.980$-$0.991$, indicating approximate normality sufficient for valid inference.

Five-fold subject-level cross-validation assessed model generalization to new individuals. Out-of-sample $R^2$ ranged from 0.22 to 0.32 (VeriEye: 0.31-0.32; OpenIris: 0.22-0.28), substantially lower than within-sample marginal $R^2$ (0.75-0.77), consistent with high ICC ($\approx$ 0.65). This gap reflects the dominant contribution of stable between-subject differences that cannot be predicted from measured covariates alone. Root mean squared error was approximately 102 points (VeriEye) and 0.040 HD units (OpenIris), approximately 0.8 SD, with consistent performance across folds.

\section{Discussion}
\label{discussion}

This study provides the most extensive longitudinal evidence to date that iris recognition remains operationally reliable throughout childhood and adolescence, with false non-match rates consistently below 0.5\% across nine years of developmental change. This performance approaches adult-level accuracy under equivalent security specifications, challenging assumptions that developmental processes necessarily compromise biometric permanence. However, our findings reveal that this stability is neither uniform across recognition algorithms nor independent of enrollment age and acquisition quality, with important implications for system design and operational policy.

The high intraclass correlation (ICC $\approx$ 0.65) indicates that approximately 65\% of performance variance reflects stable between-subject differences rather than temporal changes. This underscores that individual baseline characteristics, largely determined by initial enrollment quality and developmental maturity, dominate over aging effects in determining long-term recognition accuracy.

The most significant contribution of this work is the demonstration that apparent temporal degradation in pediatric iris recognition may reflect developmental confounding rather than genuine template aging, and that this distinction is algorithm-dependent.

VeriEye's observed performance decline was fully explained by enrollment cohort effects: children enrolled at younger ages (4-5 years) exhibited persistently lower performance than those enrolled at older ages (8+ years), with effect sizes of 0.5-0.6 standard deviations. However, these differences remained constant over time rather than worsening with elapsed years. Significant age-by-time interactions further revealed that younger enrollees exhibited steeper apparent decline trajectories, though these patterns reflect developmental variability in acquisition quality rather than accelerated biological aging. When biological age change was properly modeled, temporal effects disappeared, indicating that VeriEye's apparent degradation reflects the developmental maturity at enrollment rather than intrinsic template aging.

In contrast, OpenIris demonstrated genuine temporal aging independent of enrollment age, with Hamming distances increasing modestly but consistently over the observation period (approximately 0.5 standard deviations over eight years). This divergence likely stems from algorithmic design: OpenIris implements Daugman's binary iris code methodology, which is highly sensitive to fine-scale textural changes that may occur during development, while VeriEye's proprietary feature extraction appears more robust to such changes but correspondingly more sensitive to initial enrollment quality. This matcher-dependent behavior underscores that algorithmic choice critically influences pediatric performance and that universal conclusions about iris permanence cannot be drawn without algorithm-specific validation.

Beyond temporal effects, image quality and pupil dilation constancy emerged as the dominant predictors of matching performance, with dilation effects reaching 3.0-3.5 standard deviations for VeriEye, exceeding all temporal and developmental factors. Pupil size variability induces non-linear geometric transformations in iris texture that rubber-sheet normalization cannot fully compensate, an effect amplified in pediatric populations by greater autonomic variability and environmental sensitivity.

Failure analysis corroborated this finding: all 203 false non-matches concentrated in 26 subjects (9.4\%) with persistent acquisition challenges including blur, occlusion, and partial closure, rather than accumulating systematically with temporal separation. These results indicate that acquisition conditions, not biological template aging, represent the primary limitation in pediatric iris recognition.

This finding contrasts with other biometric modalities: face recognition exhibits substantial degradation from craniofacial growth, fingerprints show increased variability from size and elasticity changes, and voice biometrics demonstrate marked instability from vocal tract development. The iris appears comparatively stable during development, consistent with its early embryological formation, though realizing this stability requires careful attention to acquisition quality.

These findings have direct policy implications for operational biometric deployments. Current conservative re-enrollment schedules, such as India's Aadhaar program requiring updates at ages 5 and 15 or Canada's NEXUS program using five-year cycles, lack longitudinal validation and appear unnecessarily restrictive. Our nine-year stability data suggest that enrollment validity periods could be safely extended to 10-12 years for children enrolled at ages 7 or older with high-quality initial capture, reducing operational costs and user burden while maintaining security.

For very young enrollees (ages 4-6), a single re-enrollment at age 10-12 would capitalize on improved cooperation and image quality while remaining less burdensome than current multi-update schedules. Systems targeting very young children should invest in specialized acquisition protocols, trained operators skilled in pediatric interaction, and child-friendly interfaces to maximize initial enrollment quality, or alternatively consider delayed enrollment to ages 7-8 when behavioral cooperation and postural control improve substantially.

Algorithm selection should account for temporal behavior: systems prioritizing long-term stability without re-enrollment may favor algorithms exhibiting cohort-dependent but temporally stable performance like VeriEye, while those accepting modest drift may prefer alternatives like OpenIris that show genuine but operationally manageable aging. Critically, our finding that pediatric-calibrated thresholds reduced OpenIris FNMR by 89\% (from 2.65\% to 0.29\%) while maintaining equivalent security (FMR $\approx$ 0.1\%) demonstrates that default adult-optimized configurations systematically disadvantage younger users, emphasizing the importance of demographic-specific threshold calibration for achieving inclusive biometric systems.

Several limitations warrant consideration. This single-site study in Potsdam, NY, with predominantly Caucasian participants may not generalize across diverse populations, though iris recognition is generally robust to ethnic variation due to its reliance on structural rather than pigmentation-based features. The COVID-19 suspension created temporal discontinuities that may have introduced unmeasured confounds, though statistical models account for irregular spacing.

Quality and developmental age remain inherently confounded: younger children systematically produce lower-quality images due to behavioral factors, limiting our ability to fully disentangle acquisition from biological effects. Finally, this study evaluated only near-infrared imaging and verification scenarios; visible-wavelength acquisition and identification (1:N) performance may exhibit different characteristics.

Future work should address these limitations through multi-site studies across diverse populations, evaluation of mobile and visible-wavelength iris recognition, investigation of identification performance and personalized threshold adaptation, and extended follow-up into early adulthood to complete the developmental arc.

In summary, iris recognition achieves operational reliability in pediatric populations when properly implemented, with acquisition quality rather than biological aging representing the primary performance determinant. The algorithm-dependent temporal behaviors we identify (developmental confounding versus genuine aging) have significant implications for system design, policy formulation, and the broader goal of achieving demographically inclusive biometric systems. Methodologically, our age-period-cohort modeling framework addresses the fundamental identification problem inherent in longitudinal designs, enabling explicit decomposition of enrollment cohort effects, biological maturation, and elapsed time, a distinction essential for interpreting developmental biometric data but absent from prior pediatric studies. As biometric deployments increasingly serve diverse populations across the lifespan, demographic-specific validation and evidence-based policy formulation become essential to ensuring both technical performance and equitable access to identity services.

\section{Conclusion}
\label{conclusion}

This study presents the most extensive longitudinal evaluation of pediatric iris recognition to date, following 276 subjects for up to nine years through adolescence. Using an age-period-cohort modeling framework to disentangle developmental effects from temporal aging, we demonstrate that iris recognition remains operationally reliable throughout childhood, with false non-match rates below 0.5\% using pediatric-calibrated thresholds. We reveal algorithm-dependent temporal behaviors: VeriEye's apparent decline reflects developmental confounding rather than genuine template aging, while OpenIris exhibits modest but genuine aging. Image quality and pupil dilation constancy dominated performance over temporal factors. These findings support extending re-enrollment validity periods to 10-12 years for children enrolled at ages 7 or older, and highlight the necessity of algorithm-specific validation and demographic-specific calibration for equitable biometric systems.

\bibliographystyle{IEEEtran}
\bibliography{bibliography}

\end{document}